\documentstyle[prd,aps,preprint]{revtex}
\tightenlines

\def\be{\begin{equation}}
\def\ee{\end{equation}}
\def\ba{\begin{eqnarray}}
\def\ea{\end{eqnarray}}
\def\br{\begin{array}}
\def\er{\end{array}}

\def\simlt{\mathrel{\lower4pt\vbox{\lineskip=0pt\baselineskip=0pt
           \hbox{$<$}\hbox{$\sim$}}}}
\def\simgt{\mathrel{\lower4pt\vbox{\lineskip=0pt\baselineskip=0pt
           \hbox{$>$}\hbox{$\sim$}}}}
\begin{document}

\draft

\title{Large Lepton Mixings Induced by Sterile Neutrino}

\author{
K.R.S. Balaji $^{1,2}$,
A. P\'erez-Lorenzana$^{1,3}$ 
and A. Yu. Smirnov$^{1,4}$}

\address{
$^1$ The Abdus Salam International Centre for Theoretical Physics, I-34100,
Trieste, Italy\\
$^2$ The Institute of Mathematical Sciences, Taramani, Chennai 600 113, India\\
$^3$  Departamento de F\'\i sica,
Centro de Investigaci\'on y de Estudios Avanzados del I.P.N.\\
Apdo. Post. 14-740, 07000, M\'exico, D.F., M\'exico\\ 
$^4$ Institute for Nuclear Research  of Russian Academy of Sciences, Moscow
117312, Russia }

\date{March, 2000}

\maketitle

\begin{abstract} 

We assume that the mass matrix of active neutrinos has hierarchical form with
small mixings, similar to quark mass matrix. We show that the large mixings
between $\nu_\mu$ and  $\nu_\tau$ as well as between   $\nu_e$ and certain
combination of  $\nu_\mu$ and $\nu_\tau$    required by the present data can 
appear due to the presence of a sterile neutrino state. Two  realizations of
this possibility are considered:  (i) Large flavor mixing appears as a result
of sterile neutrino ``decoupling''  ($m_{ss} \gg m_a$), so that the
active-sterile mixings are negligible. (ii) Sterile neutrino has a mass of
$O(1)$ eV and its mixing with active neutrinos can be observable. In the
second  case,  the (3+1) scheme of neutrino mass, which also accommodates  the
LSND  result, can be reproduced, provided the  hierarchy of the  mass matrix of
active neutrinos is not strong (the ratio of  largest to smallest elements is
about two orders of magnitude).
The enhancement of lepton mixing via coupling with
sterile neutrino  can be realized in Grand Unified theories with  the see-saw
mechanism of neutrino mass generation. \\[1ex]
\end{abstract}

\section{Introduction}
 
 With high confidence levels, one can conclude that
 the atmospheric neutrino data~\cite{ref1} imply
large or even maximal mixing of the muon neutrino. Recent detailed studies
indicate that $\nu_\mu$ strongly mixes with  $\nu_\tau$ while  
mixing with the sterile neutrino $\nu_s$ is disfavoured at the
$3\sigma$ level~\cite{ref2}. The best 
global fit of the solar neutrino data is given
by the LMA (large mixing angle) solution, although small mixing angle 
solution (SMA) is not ruled out~\cite{ref3}. This striking difference of 
large lepton mixing and a small
quark mixing raises important questions about the origins of neutrino masses, 
quark-lepton symmetry and Grand Unification. 

There has been a lot of efforts 
to explain the appearance of large lepton mixing and small quark mixing in
an unique framework.
Within the context of see-saw mechanism of neutrino mass 
generation~\cite{ref4},
which realizes quark-lepton symmetry most closely, large lepton mixing
can appear due to large mixing in the charge lepton mass 
matrix~\cite{ref5} 
or in Dirac mass matrix of neutrinos~\cite{ref6,king}. It can also appear
from 
large mixing or very strong hierarchy in the Majorana mass matrix of the 
right handed neutrinos~\cite{ref7}. 
Large lepton mixing can be obtained by radiative corrections
due to renormalization group effects in  schemes with quasi-degenerate 
neutrinos~\cite{ref8}. A number of  possibilities exist, where the neutrino 
masses and the masses of the charged fermions have different origins,
e.g., small neutrino masses appear due to loop effects~\cite{ref9}. 

Notice that, the mechanisms suggested so far imply an explicit
introduction 
of large mixing in one of the sectors of the theory, or a very
strong mass hierarchy or a strong mass degeneracy or qualitatively new 
mechanism of neutrino mass generation~\cite{ref10}.

Another possible difference between the quark and lepton sectors is that 
more than three light neutrinos may be involved in lepton mixing.
Introduction of an
additional sterile  neutrino state is motivated by simultaneous explanation of
the solar, atmospheric and LSND data~\cite{ref11} in terms of neutrino
oscillations~\cite{ref12}. 

In this paper we address the question: 
can these two issues, large mixing of the active neutrinos 
(flavor mixing) 
and an existence of sterile neutrino be related? We suggest  
a mechanism of enhancement of the flavor mixing, which is related to the 
existence of a sterile neutrino.

One remark is in order. It seems there is no simple relations between
masses and mixing both in quark and in  lepton sectors. This may 
be an indication  of complex origin of the masses 
when two or more different mechanisms give comparable contributions.  

The paper is organized as follows. In the next section 
we formulate and discuss our context. In sect. III, we 
describe a mechanism of enhancement of the mixing via decoupling of 
sterile neutrino. In Sect. IV we consider the enhancement in the
non-decoupling limit and show how the LSND result can be accommodated.   
In sect. V we argue that the suggested mechanism can be consistent with 
the see-saw mechanism and  Grand Unification.

\section{Flavor matrix and  $\nu_s$-mixing}

We  assume that the sterile neutrino, $\nu_s$,
exists and mixes with the active neutrinos  
$\nu_{\alpha}$, $(\alpha = e , \mu, \tau)$.  
In the basis $(\nu_e, \nu_{\mu}, \nu_{\tau}, \nu_s)$
the mass matrix can be written as
 \be 
 \hat M = \left( \begin{array}{c c}
           \hat{m}_f &  {\vec\epsilon}m_{ss}\\
	{\vec\epsilon}~^\dagger m_{ss} & m_{ss}
       \end{array} \right) ~,
  \label{eq1}
  \ee
where $\hat{m}_f$ is the $3\times3$ mass matrix of the active neutrinos
which we will call the flavor matrix, $m_{ss}$ is the Majorana mass term
for the sterile neutrino and the vector 
${\vec\epsilon}= (\epsilon_e,\epsilon_\mu,\epsilon_\tau)$ 
parametrizes the active-sterile mixing masses  in units of $m_{ss}$.
The parameters $\epsilon_{\alpha}$ determine the mixing between the
active and 
sterile neutrinos. 

We assume that matrix $\hat m_f$ has a hierarchical structure with small
mixings, similar to the quark mass matrix. 
Such a  matrix is usually generated 
by the see-saw mechanism in the context of Grand Unified theories 
or in general, in the theories 
with quark-lepton symmetry ~\cite{models}. Indeed, in the GUT   
the Dirac neutrino mass matrix  
can be similar to the up-quark mass matrix and if there are no strong
mixings or hierarchy in  the 
Majorana mass matrix of the right handed neutrinos, $\hat M_{R}$, 
the see-saw mechanism, 
$\hat{m}_f \approx -  \hat m_D \hat M^{-1}_{R} \hat m_D^{\dagger}$,  can 
produce $\hat{m}_f$, which  has a structure similar to $\hat m_D$
\cite{ref10}. Obviously, $\hat{m}_f \propto \hat m_D$, if 
$\hat M_{R} \propto \hat m_D$.  Certain  realistic mass matrices 
$\hat m_D$ satisfy the condition $m_D \cdot m_D^{\dagger} \propto m_D$,
so that original  structure of $\hat m_D$ is reproduced by the product. 
In this case $\hat{m}_f \propto \hat m_D$, if 
$\hat M_{R} \propto \hat I$, where 
$\hat I$ is the  unit matrix (that is,  the matrix is flavor blind). 
In a number of cases the see-saw mechanism leads to milder hierarchy of 
$\hat{m}_f$ (factor 2 - 3 enhancement of small elements). However, only in
extreme cases  near maximal mixing appears \cite{ref7,kuo}.  

In what follows, we will show 
that the mixing of the active neutrinos with the sterile state can  produce 
large mixings of  the active neutrinos. We will consider two possibilities 
(two limits) which have  different phenomenological
consequences: (i) the decoupling limit of the sterile neutrino:  
when $m_{ss} \gg m_{\tau\tau}$ and $\nu_{\alpha} - \nu_s$ mixing 
is very small ($\epsilon \ll 1$); (ii) non-decoupling limit 
when $m_{ss} \simgt m_{\tau\tau}$ and $\epsilon_{\alpha}$  are not too small,
so that $\nu_{\alpha} - \nu_s$ mixing can be  observable.

\section{Large mixing via decoupling}

Let us assume that $\epsilon_{\alpha} \ll 1$ and  
 $m_{ss} \gg m_{\alpha \beta}$, 
where  $m_{\alpha\beta}$ are the elements of the flavor  matrix
$\hat m_f$. After decoupling 
of $\nu_s$ which is equivalent to the block diagonalization of the  
matrix (\ref{eq1})  we get the flavor matrix $m_f^\prime$ with elements 
\be
m_{\alpha\beta}^\prime  \cong m_{\alpha\beta} -
\epsilon_\alpha \epsilon_\beta m_{ss}~, 
\label{eq3}
\ee
where the  second term, 
$\epsilon_{\alpha}\epsilon_{\beta} m_{ss} \equiv
\Delta m_{\alpha \beta}$, is the contribution from 
the $\nu_s$-mixing. 
We define the decoupling limit in such a way that corrections 
to (\ref{eq3}) being of the order $O(\epsilon^3 m_{ss})$, are negligible. 

We require that matrix $\hat m_f^\prime$ leads to oscillation parameters 
which can explain the atmospheric and the solar neutrino data. 
That is, the eigenvalues, $m_i$ $(i = 1, 2, 3)$ and the mixing angles 
should satisfy the following condition:  
$m_3^2 - m_2^2 = \Delta m^2_A \sim (2 - 6)\cdot 10^{-3}$ eV$^2$,  
$\sin^2 2\theta_{\mu\tau} = \sin^2 2\theta_A  \gtrsim 0.88$, 
$m_2^2 - m_1^2 \equiv  \Delta m^2_{\odot}$ and 
$\sin^2 2\theta_{e x} = \sin^2 2\theta_{\odot}$.  
Here $x$ refers to some combination of $\nu_{\mu}$ and $\nu_{\tau}$ and 
$\Delta m^2_{\odot}$,  $\theta_{\odot}$ are the solar oscillation
parameters from  one of the region of solutions 
for the solar neutrino problem. 

In the case of mass hierarchy: 
$m_3 \approx \sqrt{\Delta m^2_A} \gg m_2 \sim \sqrt{\Delta m^2_{\odot}}$, 
the requirement of the near to maximal mixing of 
$\nu_{\mu}$ and $\nu_{\tau}$ reduces  to the following relation between
the matrix elements: 
\be
m_{\mu\mu}' \approx m_{\mu\tau}' \approx m_{\tau\tau}^\prime 
\approx \frac{1}{2}\sqrt{\Delta m^2_A} \equiv m_A~, 
\label{rel-matr}
\ee
where we will call $m_A$ the atmospheric mass scale. 

Diagonalizing the $\nu_{\mu} - \nu_{\tau}$ submatrix of the mass matrix 
$m_f'$   (the so called  dominant block),  
we get the off-diagonal elements 
\ba
\tilde{m}_{e \mu} &=&
\cos \theta_A m_{e \mu}' - \sin \theta_A m_{e \tau}'~,\nonumber\\
\tilde{m}_{e \tau} &=& 
\cos \theta_A m_{e \tau}' + \sin \theta_A m_{e \mu}'~, 
\label{off-diag}
\ea
where $\theta_A$ is the angle responsible for oscillations of the
atmospheric neutrinos, so that $\sin \theta_A \approx \cos \theta_A
\approx 
1/\sqrt{2}$. The element $\tilde{m}_{e \mu}$ is immediately related to
the solar neutrino oscillation parameters (under the assumption that
${m}_{e e}$ is very small): 
\be
\tilde{m}_{e \mu} \approx 
\frac{\sqrt{\Delta m^2_{\odot}}}{2}
\left[
\frac{\tan^2 2\theta_{\odot}}
{\sqrt{1 + \tan^2 2\theta_{\odot}}}
\right]^{1/2}~.
\label{m-emu}
\ee
For LMA solution, we need to have typically 
$\tilde{m}_{e \mu} = (2 - 10) \cdot 10^{-3} {\rm eV}$,  
for SMA:  
$\tilde{m}_{e \mu}$ $ = (0.3 - 1.5) \cdot 10^{-4} {\rm eV}$, and for 
LOW, $\tilde{m}_{e \mu}$ $ = (1 - 2) \cdot 10^{-4} {\rm eV}$.

The mass  $\tilde{m}_{e \tau}$ is related to the $U_{e3}$ element of 
mixing matrix by 
\be
\tilde{m}_{e \tau} \approx \sqrt{\Delta m^2_A} U_{e3}~,
\label{m-etau}
\ee
which in turn is restricted by the CHOOZ result, 
$U_{e3} \lesssim 0.1$, so that 
$
\tilde{m}_{e \tau} \lesssim (4 - 8) \cdot 10^{-3} {\rm eV}~. 
$

Phenomenological requirements (\ref{m-emu},\ref{m-etau}) and, 
especially (\ref{rel-matr}),  
can not be satisfied  for arbitrary original matrix $\hat m_f$. 
However, as we will show,  they can be satisfied due to 
$\nu_s$-mixing effect for a wide class of rather plausible 
matrices $\hat m_f$ with hierarchy and small mixing.  Parameters of 
the $\nu_s$- mixing described by $\epsilon_{\alpha}$ depend on 
structure of  the  matrix $\hat m_f$. Here, we consider three 
examples which differ by scale of masses in $\hat m_f$.

\subsection{Large masses in $\hat m_f$} 

Suppose that 
\be
m_{\tau \tau} \gg m_{\tau \mu} \gg m_{\mu \mu} \sim m_A~, 
\label{first}
\ee
(two largest elements in $\hat m_f$ are much heavier than the
atmospheric mass scale). For instance, the elements of   
$\hat m_f$ may have the following hierarchical structure: 
$m_{\tau \tau} = m_0 \sim O(10~ {\rm eV})$,   
$m_{\tau \mu} \sim m_0 \lambda^2$, 
$m_{\mu \mu} \sim  m_{e \tau} \sim m_0 \lambda^4$,    
$m_{e \mu} \sim m_0 \lambda^6$, 
$m_{e e} \sim m_0 \lambda^8$, 
where $\lambda \sim 0.2$, is the hierarchy parameter of the order of
Cabibbo angle. 

The contribution from the $\nu_s$-mixing should lead to strong
cancellation in $m_{\tau \tau}'$ and   $m_{\tau \mu}'$. 
{}From the condition $m_{\tau \tau}' \sim m_{\tau \mu}' \sim m_A$ we get 
\be
\epsilon_{\tau} \cong \sqrt{\frac{m_{\tau \tau}}{m_{ss}}}, ~~~ 
\frac{\epsilon_{\mu}}{\epsilon_{\tau}} \cong  
\frac{m_{\tau \mu}}{m_{\tau \tau}} \sim \lambda^2~. 
\label{eps1}
\ee
Furthermore, using (\ref{eps1}) we find  
for the element $m_{\mu \mu}'$ 
\be
m_{\mu \mu}' = m_{\mu \mu} - \frac{m_{\tau \mu}^2}{m_{\tau \tau}} 
\sim m_A~, 
\label{m-mumu}
\ee
and in fact,  $m_{\mu \mu}$ can be much smaller than $m_A$.  Notice that the
signs of $m_{\mu \mu}'$ and $m_{\tau \tau}'$ should be the same, which can be
realized if $|m_{\tau \tau}| <|\epsilon_{\tau}^2m_{ss}|$
 
In our example of $m_f(\lambda)$, we get 
$m_{e \mu} \ll m_{e \tau} \sim 10^{-2}$ eV, so that 
$|\tilde{m}_{e \mu}| \approx |\tilde{m}_{e \tau}| \sim  
5 \cdot 10^{-3}$ eV. These values  satisfy the CHOOZ bound 
and lead to the LMA solution. (The expected value 
of $U_{e3}$ is close to the upper bound.)  
Therefore, no significant contribution 
of the $\nu_s$-mixing to these elements 
is required, 
and we can set an upper bound on $\epsilon_e$. From the condition 
$\Delta m_{e \tau} =$  
$(\epsilon_e/\epsilon_{\tau})m_{\tau \tau}  \simlt 10^{-2}$ eV we get 
\be
\frac{\epsilon_e}{\epsilon_{\tau}} \simlt 10^{-3}.
\label{h-eps}
\ee
Actually, the $\nu_s$-mixing with $\epsilon_e/\epsilon_{\tau} \sim
10^{-3}$  
can be used to suppress $U_{e3}$ and/or to reproduce  the SMA solution. 

Thus, the hierarchical structure of $\nu_s$-mixing  is required 
$\epsilon_e : \epsilon_{\mu} : \epsilon_{\tau} =  (\simlt 10^{-3}): 0.04: 1$
(in our example) which should correlate with hierarchy of the original
mass matrix. 
This can be a consequence of certain flavor symmetry. 

\subsection{Moderate masses in $\hat m_f$} 

Suppose that
\be
m_{\tau \tau} \gg m_{\tau \mu} \sim m_A~, 
\label{moderate}
\ee
(only $m_{\tau \tau}$ is much larger that the atmospheric neutrino
scale). As we will see, the contribution of 
the $\nu_s$-mixing to $m_{\mu \mu}$ is small, so that 
 the condition (\ref{rel-matr}) can be satisfied only if 
\be
m_{\mu \mu} \sim m_A. 
\ee

The original flavor matrix $\hat m_f$  could have the following
structure: 
$m_{\tau \tau} = m_0 \sim O(1 ~{\rm eV})$,
$m_{\tau \mu} \sim   m_{\mu \mu} \sim m_0 \lambda^2$,
$m_{e \mu} \sim  m_{e \tau} \sim m_0 \lambda^4$,
$m_{e e} \sim m_0 \lambda^6$. 

{}From the condition of cancellation in $m_{\tau \tau}'$ we get 
for $\epsilon_{\tau}$ the same relation as in 
(\ref{eps1}). Contribution of the $\nu_s$-mixing to $m_{\tau \mu}$ 
should not be large. Requiring  
$\Delta m_{\tau \mu} \simlt m_A$, we find 
\be
\frac{\epsilon_{\mu}}{\epsilon_{\tau}} \lesssim 
\frac{\sqrt{\Delta m^2_A}}{2m_{\tau \tau}} = 
(2 - 4)\cdot 10^{-2}\left(\frac{\rm 1~ eV }{m_{\tau \tau}}\right)~.  
\label{eps-mod}
\ee
One can check now that the correction to $m_{\mu \mu}$ is indeed small:  
$\Delta m_{\mu \mu} \sim 
(\epsilon_{\mu}/\epsilon_{\tau})^2 m_{\tau \tau} 
= (0.4 - 1.6) \cdot 10^{-3} \ll m_A$. 

For our example of $m_f(\lambda = 0.2)$ we get 
$m_{e \mu} \sim m_{e \tau} \sim (1 - 2) \cdot 10^{-3}$ eV, so that
$|\tilde{m}_{e \mu}| \approx |\tilde{m}_{e \tau}| \sim
(0 - 3) \cdot 10^{-3}$ eV. 
Therefore the CHOOZ bound is satisfied, and moreover,    
expected  $U_{e3}$ is much below the upper bound.
Any solution for the solar neutrino problem can be reproduced 
depending on precise values of parameters in $\hat m_f$. 
Therefore, no significant contribution to these elements
from the $\nu_s$-mixing is required. 
The  upper bound on $\epsilon_e$ can be  obtained 
from the condition, $\Delta m_{e \tau}  \leq 10^{-2}$ eV: 
$(\epsilon_e/\epsilon_{\tau}) \leq 10^{-2}$. 
Thus, $\epsilon_{\alpha}$ are hierarchical, 
$\epsilon_e : \epsilon_{\mu} : \epsilon_{\tau} = 
(< 10^{-2}): 0.04: 1$,although the hierarchy can be  weaker 
than  in the previous case.

Let us consider another  structure of the original 
flavor mass matrix:  $m_{e \mu} \sim  m_{e \tau} \sim m_0 \lambda^3$,
$m_{e e} \sim m_0 \lambda^4$. In this case, 
$m_{e \mu} \sim m_{e \tau} \sim (4 - 8) \cdot 10^{-3}$ eV, 
which can give  
$|U_{e3}|$ at the level of the CHOOZ bound 
and lead to the LMA solution  without any contribution from the 
$\nu_s$-mixing. 
For $\epsilon_e \sim 10^{-2}\epsilon_{\tau}$,  
cancellation between $m_{e \mu}'$ and $m_{e \tau}'$ 
in $\tilde{m}_{e \mu}$ allows  one to 
reproduce also the SMA solution. 

\subsection{Small masses in $\hat m_f$}
 
Suppose that
\be
m_{\mu \mu} \ll m_{\tau \mu} \ll m_{\tau \tau}  \sim m_A~,
\label{small}
\ee
i.e., all the masses in the original flavor matrix 
are smaller than the atmospheric mass scale. 
As an example we can take   
$m_{\tau \tau} = m_0 \sim 0.03~ {\rm eV}$,
$m_{\tau \mu} \sim m_0 \lambda$, 
$m_{\mu \mu} \sim m_{e \tau} \sim m_0 \lambda^2$, 
$m_{e \mu} \sim m_0 \lambda^3$, 
$m_{e e} \sim m_0 \lambda^4$. In this case, the entire 
dominant block is formed by contributions 
from the $\nu_s$-mixing and the original flavor matrix 
gives small corrections. In order to satisfy  
(\ref{rel-matr}) we should take $\epsilon_{\mu} \sim \epsilon_{\tau}$. 
Neglecting all mass terms but $m_{\tau \tau}$ in $\hat m_f$, we have 
\ba
&&
m_{\mu \mu}' \approx - \epsilon_{\mu}^2 m_{ss}~;~ 
m_{\mu \tau}' \approx - \epsilon_{\mu} \epsilon_{\tau}  
m_{ss}~;\nonumber\\
&& m_{\tau \tau}' \approx m_{\tau \tau} - 
\epsilon_{\tau}^2 m_{ss}~.  
\label{mass-small}
\ea
Choosing $\epsilon_{\tau}^2 m_{ss} = 2 m_{\tau \tau}$, 
from the  condition $|m_{\tau \tau}'| \sim  |m_{\tau \mu}'|\sim m_A$   
we get    $\epsilon_{\mu}/\epsilon_{\tau} = 1/2$.

The  mass $m_2$ responsible for the solar neutrino conversion
is determined by $m_{\tau \tau}$. The determinant of the 
$\nu_{\mu} -  \nu_{\tau} $ submatrix equals 
$- \epsilon_{\mu}^2 \epsilon_{\tau}^2  m_{ss}^2$ 
$ - m_{\mu \mu}' m_{\tau \tau} \approx m_A m_{\tau \tau}/k$, where 
$k = O(1)$ is a free parameter. 
Therefore, taking into account that $m_3 \sim 2 m_A$, we get 
\be
m_2 \sim \frac{m_{\tau \tau}}{2k} \sim \frac{\sqrt{\Delta m^2_A}}{4k}~. 
\label{m-2}
\ee
For $k =2$ we find the mass $m_2 = (0.5 - 1) \cdot 10^{-2}$ eV, which is
well in the 
range of the LMA solution, furthermore $\sin^2 2\theta_A = 0.94$. 

For other  elements (without $\nu_s$-mixing contribution) we have, 
$m_{e \mu} \ll  m_{e \tau} \sim (1 - 2) \cdot 10^{-3}$ eV, so that
$|\tilde{m}_{e \mu}| \approx |\tilde{m}_{e \tau}| \sim
(0.5 - 1) \cdot 10^{-3}$ eV, which is much smaller than  value required by
the LMA solution. However, 
$m_{e \mu}$ can be enhanced by the $\nu_s$-mixing contribution. 
Requiring that $\Delta m_{e \tau} \sim 6 \cdot 10^{-3}$ eV, we get 
$\epsilon_e/ \epsilon_{\tau} \sim 1/10$. Thus the scheme implies 
a rather weak  hierarchy of mixing parameters: $\epsilon_e :
\epsilon_{\mu} : \epsilon_{\tau} = 
0.1 : 0.5 : 1$. 

\subsection{On the scale of $m_{ss}$} 

Clearly, all 
the contributions from the $\nu_s$-mixing  will be unchanged if 
$\epsilon_{\alpha}$  decreases  with increase of 
$m_{ss}$ as 
\be
\epsilon_{\alpha} =  
\epsilon_{\alpha}^0 \sqrt{\frac{m_{ss}^0}{m_{ss}}}~.
\label{scale}
\ee

In our examples the largest $\epsilon_{\alpha}$  
is  $\epsilon_{\tau} = \sqrt{m_{\tau \tau}/m_{ss}}$, and 
the decoupling limit holds if $\epsilon_{\tau} < (10^{-1} - 10^{-2})$ 
depending on the type  of the flavor mass matrix. 
Let us consider several possibilities. 

For $m_{\tau \tau} \sim 1$ eV, the decoupling implies that 
$m_{ss}$ should be at least in the kev range. 
The kev  sterile neutrinos with even very small mixing 
(as small as $\epsilon \sim 10^{-2} - 10^{-4}$)
may have important consequences both in astrophysics and cosmology.  
They have been discussed as a possible component of the 
(warm) dark matter of the universe~\cite{Juha},  in connection 
with pulsar kick problem \cite{kusenko,nardi}, the proton loading problem    
for gamma bursters \cite{gamma}. 
Such neutrinos can be  produced in the  conversion 
of active neutrinos  
in the supernovae and in the early Universe. 
To satisfy the cosmological bound these neutrinos 
should decay during the time shorter than the 
lifetime of the universe or one should assume  that these neutrinos 
as well as active neutrinos were never in thermal equilibrium 
\cite{riotto}.

If $m_{ss} \sim 1$ TeV, which one could  expect 
for singlets in the supergravity models (modulinos) \cite{roulet} 
then $\epsilon_{\tau} \sim 10^{-6}$,  so that the mixing terms 
$\epsilon_{\tau}m_{ss}$ should be  about 1 MeV. 
For $m_{ss} \sim 10^{13}$ GeV, we have 
$\epsilon_{\tau} \sim 10^{-11}$ and $\epsilon_{\tau}m_{ss} \sim100$ 
GeV, which corresponds to the electroweak  scale.

\section{Non-Decoupling Limit and LSND result}

\subsection{Without LSND result}

Here we will assume that $m_{ss} \sim (1 - 10)$ eV. In the case 
of strong hierarchy of the mixing parameters 
$\epsilon_{\tau} \gg \epsilon_{\mu} \gg \epsilon_e$ as in 
examples 1) and 2) from the previous section the non-decoupling effect
reduces  to relatively large mixing 
of the $\nu_{\tau}$ and $\nu_s$ in the fourth mass eigenstate. 
So, in principle,   one can observe $\nu_{\tau} \leftrightarrow \nu_s$ 
oscillations driven by $\Delta m^2 \sim m_{ss}^2$
in the short base-line experiments. 
All other masses and mixings will be as in the decoupling limit. 

Let us now discuss a possibility to realize the (3 + 1) scheme 
\cite{olg} with  
relatively large  $\epsilon_e$ and $\epsilon_{\mu}$ which 
can also accommodate the LSND result. Now block diagonalization is
impossible and we need to consider whole $4\times4$ matrix 
$\hat M$. 

First of all, it is easy to see that if
the flavor matrix $\hat m_f$, has a strong hierarchy, the LSND 
result can not be reproduced because of 
the CHOOZ bound. Let us 
consider the matrix $\hat m_f$ with only one 
non-zero element, $m_{\tau\tau}$. As follows from 
(\ref{eq1}),  in this $m_{\tau\tau}$ dominance approximation  
one state 
\be
\nu_d \equiv \cos \phi ~\nu_e - \sin \phi ~\nu_\mu~,
\label{s2}
\ee
where $\cos \phi  = \epsilon_\mu/\sqrt{\epsilon_e^2 + \epsilon_\mu^2}$ 
decouples and the orthogonal state 
\be
\nu_n \equiv \sin \phi ~\nu_e + \cos \phi ~\nu_\mu 
\label{s3}
\ee
will have the mixing element, $\sqrt{\epsilon_e^2 + \epsilon_\mu^2} m_{ss}$ 
in the rest $3\times 3$ matrix.
To explain the atmospheric neutrino data, one needs to have large   
admixture of $\nu_n$ (which contains $\nu_\mu$) in the $\nu_3$ - state.  
Denoting by $U_{n3}$~, this admixture we get
\be
U_{\mu3} = \cos \phi \cdot  U_{n3},~~~~ U_{e3} = \sin \phi \cdot U_{n3}.
\label{s4}
\ee
{}From (\ref{s4}), we get
\be
U_{e3}=U_{\mu3}\cdot \tan \phi = U_{\mu3}\cdot 
\frac{\epsilon_e}{\epsilon_\mu}~,
\label{s5}
\ee
where the atmospheric data imply $U_{\mu 3}\simgt 0.5$. The LSND result, 
in turn, requires  $\epsilon_e \sim 0.12 - 0.14$ and  $\epsilon_\mu \sim
0.12 - 0.16$,
so that  from (\ref{s5}) we get  $|U_{e3}|^2 > 0.12$ 
which is certainly excluded by
CHOOZ result~\cite{chooz}. 
Thus, large mixing of $\nu_\mu$ in $\nu_3$ leads simultaneously to large
mixing of $\nu_e$ in $\nu_3$.

Taking $\epsilon_e/\epsilon_\mu \sim 1/3$ to statisfy
the CHOOZ limit, we get for $\epsilon_\mu \leq 0.15$, 
$\sin^2 2\theta_{LSND} \approx 4 \epsilon_e^2 \epsilon_\mu^2 \leq 2.5 \cdot
10^{-4}$, which is smaller than the experimental result~\cite{lsnd}. This 
requires a departure from the $m_{\tau\tau}$ dominance scheme.

\subsection{With LSND result}

In what follows, we show that the LSND result can be reproduced if  
the flavor matrix has a moderate hierarchy: 
\be
M_{\alpha \beta} \sim \frac{1}{3} M_{\beta \beta}~, 
\label{hier}
\ee
where $\alpha$ corresponds to lighter flavor than $\beta$. 

The mass matrix in the flavor basis can be written as 
\be
\hat M = \hat U\hat M^{diag}\hat U^{\dagger}~, 
\label{matr}
\ee
where $\hat U$ is the mixing matrix, and the diagonal matrix of the mass
eigenvalues, $\hat M^{diag}$, can be parametrized as 
\be
\hat M^{diag} = m_1 \hat{I} + diag(0, \delta, 0, 0) + diag(0, 0, \Delta,
m),   
\label{diag}
\ee
so that $m_2 = m_1 + \delta$, $m_3 = m_1 + \Delta$, and 
$m_4 = m_1 + m$. Inserting (\ref{diag}) into (\ref{matr}) 
we get 
\be
\hat M = m_1 \hat{I} + \hat M^{\delta} + 
\hat U diag(0, 0, \Delta,  m)\hat U^{\dagger},
\label{matr2}
\ee   
where $\hat M^{\delta} \equiv \hat U diag(0, \delta, 0,  0)
\hat U^{\dagger}$.  

Let us first assume that $m_1 \gg \delta$, so that two light states 
are degenerate, $m_1 \approx m_2$. Neglecting $\hat M^{\delta}$, 
we get for  the elements of matrix $\hat M$:
\be
M_{\alpha \beta} = m_1 \delta_{\alpha \beta} + 
\Delta U_{\alpha 3} U^{\dagger}_{\beta 3} + 
m U_{\alpha 4} U^{\dagger}_{\beta 4}~. 
\label{element}
\ee
That is, the structure of the mass matrix is determined by
the mass of the degenerate pair, $m_1$, by mass parameters 
$\Delta$ and $m$ and by flavour mixing in the
third and in the fourth mass eigenstates $U_{\alpha 3}$, $U_{\alpha 4}$.
These mixing elements 
are, in turn,  determined  by phenomenology and
a condition of the mass hierarchy. Indeed,  the LSND result
requires admixtures of the $\nu_e$  and $\nu_{\mu}$ in 
$\nu_4$ to be at the level of upper experimental bounds: 
$U_{e 4} \sim  0.10 - 0.14$~\cite{bugey}, $U_{\mu 4} \sim  0.12 - 0.15$
\cite{cdhs}. 
The element $U_{e 3}$ is restricted by the CHOOZ result: 
$U_{e 3} < 0.1$ \cite{chooz}; $U_{\mu 3}$ is given by the atmospheric
neutrino 
data: $U_{\mu 3} = (0.55 - 0.85)$. One mixing parameter 
(we take $U_{\tau 3}$) is free, and  all other  parameters 
are determined by the unitarity condition (orthogonality 
of $\nu_3$ and $\nu_4$ and normalization: 
$\sum |U_{\alpha 3}|^2 = \sum |U_{\alpha 4}|^2 = 1$). 

Three lightest elements of the flavor mass matrix can then be written as 
\ba
&&
M_{ee} \approx m_1 + m |U_{e4}|^2 \approx 0.014 m + m_1~,\nonumber\\
&&
M_{e \mu}  \approx m U_{e4}  U_{\mu 4}^{\dagger} 
\approx 0.017m~,\nonumber\\
&&
M_{\mu \mu } \approx m_1 + \Delta |U_{\mu 3}|^2 + m |U_{\mu 4}|^2~,~~~~~  
\label{light}
\ea
where the contributions of terms proportional to $\Delta$ ($\Delta$- terms) 
to $M_{ee}$ and  $M_{e \mu}$ can be safely neglected. 
Clearly, to achieve the  hierarchy, 
$M_{ee} \ll M_{e \mu} \ll M_{\mu \mu }$, 
one needs significant cancellation in $M_{ee}$, and therefore
\be
m_1 = - (0.01 - 0.02) m \sim - (0.01 \div 0.02)~ {\rm eV}. 
\label{m-one}
\ee
Furthermore, $\Delta$ - term in (\ref{light}) should give  a significant
contribution to $M_{\mu \mu }$: for $\Delta = 7 \cdot 10^{-2}$ eV and 
$|U_{\mu 3}|^2 \approx 0.5$ we get $M_{\mu \mu } \sim 0.05 m$,    
which is about factor of three larger than $M_{e \mu }$. 

For $M_{\tau \tau }$ - element, we have 
\be
M_{\tau \tau}  \approx  m_1 + 
\Delta |U_{\tau 3}|^2 + m |U_{\tau 4}|^2~.
\label{m-tau}
\ee
Hierarchy between the diagonal elements, 
$M_{\tau \tau} \gg M_{\mu \mu}$, implies that the third term in 
(\ref{m-tau}) should dominate over others. 
This means that the admixture  of $\nu_{\tau}$ in  the 
fourth state should not be small. From the condition 
$M_{\tau \tau} \geq (5 - 10) M_{\mu \mu}$, we get 
\be
|U_{\tau 4}| \geq 0.4 - 0.7~. 
\label{tau-4}
\ee

Dominance of the terms proportional to $m$ ($m$- terms) 
in the elements $M_{\alpha \tau}$ leads to relations
\be
M_{e \tau} :  M_{\mu \tau} :  M_{\tau \tau} = 
|U_{e 4}| : |U_{\mu 4}| : |U_{\tau 4}|.
\label{rel-tau}
\ee
As an example we may have $M_{e \tau} :  M_{\mu \tau} :  M_{\tau \tau}
= 0.20 : 0.25 : 1$.  
For  values of mixing (\ref{tau-4}) 
we get, $M_{\tau \tau} = (0.17 - 0.45)$ eV 
(if $m = 1$ eV). 

We evaluate the effect of matrix $M^{\delta}$ 
in (\ref{matr2}). The parameter $\delta$, splits the light 
degenerate mass eigenvalues and, the size of $\delta$ is determined by the 
solar mass splitting $\Delta m^2_{\odot}$. For 
$m_1 = 10^{-2}$ eV and $\Delta m^2_{\odot} = (4 - 6) \cdot 10^{-5}$ 
eV$^2$ (which is typical for LMA solution) 
we get 
\be
\delta \approx \frac{\Delta m^2_{\odot}}{2 m_1} 
\approx (2 - 3) \cdot 10^{-3} {\rm eV}~.
\label{delta}
\ee
For all other solutions  value of $\delta$  and consequently the   
corrections are smaller. Given, $M^{\delta}_{\alpha \beta} = 
\delta \cdot U_{\alpha 2}  U_{\beta 2}^{\dagger}$, the largest 
mass term equals $M^{\delta} \leq 0.5 \delta \sim (1 - 2)\cdot 10^{-3}$ 
eV. Thus, the corrections are small even for the 
smallest element:  $M_{ee}$. For the remaining elements, the corrections 
are negligible and can not influence the structure of the mass 
matrix. These small corrections are, however,  important for fixing the
oscillation parameters relevant for solar neutrinos.

Let us comment on  the case of strong  hierarchy of the mass eigenvalues. 
This would correspond to the limit $m_1 \ll \delta$ and 
\be
\delta \approx \sqrt{\Delta m^2_{\odot}} \approx (5 - 10)\cdot 10^{-3} 
{\rm eV}~.
\label{delta-2}
\ee
Now the $M_{ee}$-element of the mass matrix equals  
\be
M_{ee} \approx \delta |U_{e2}|^2 + m |U_{e4}|^2~, 
\label{ee-element}
\ee
and according to (\ref{delta-2}) the first term in (\ref{ee-element}) is 
about  
$(2 - 5)\cdot 10^{-3}$ eV. It can not compensate the second term that 
is required by the hierarchy $M_{ee} \ll M_{e \mu}$. The 
compensation can be achieved if $\delta \geq 2 \cdot 10^{-2}$ eV 
which would correspond to 
$\Delta m^2_{\odot} \geq  4 \cdot 10^{-4}$ eV$^2$. This 
value is disfavored by present data, although not 
excluded completely.  

Collecting all the information discussed above we get 
as an example the following mass matrix  
\be
\hat M \approx m 
  \left(\begin{array}{c c c c}
 0.004 & 0.0168 & 0.081 & 0.14\\
 0.0168 & 0.05 & 0.13 & 0.142\\
 0.081 & 0.13 & 0.385 & 0.6\\
 0.14 & 0.142 & 0.6 & 1
\end{array} \right)~.
\label{example}
\ee
For $m = 1$eV this matrix leads to 
$\sin^2 2\theta_\odot = 0.92~;~\Delta m^2_\odot= 4.6 \cdot 10^{-5}$ eV$^2$ ; 
$\sin^2 2\theta_A = 0.958~;
~\Delta m^2_A = 3.1 \cdot 10^{-3}$ eV$^2$ ;  
$\sin^2 2\theta_{LSND} = 1\cdot 10^{-3}$.  
We also get $U_{e3}=0.099$. 
To illustrate the role of the sterile neutrino 
couplings we find that in their absence the active part
in Eq. (\ref{example}) alone  provides 
for atmospheric neutrinos: 
$~\Delta m^2 = 0.19$ eV$^2$ and  $\sin^2 2\theta = 0.35$, 
which are far beyond the required values,
and for solar neutrinos:
$~\Delta m^2 = 10^{-4}$ eV$^2$ and  $\sin^2 2\theta = 0.55$.

Thus, the  (3 + 1) scheme can be realized if the 
flavor mass matrix has a moderate hierarchy (\ref{hier}). 
There are two 
generic consequences  of the scheme: (i) the  pair of the light states 
should be degenerate with $m_1 \approx m_2 \approx (0.01 - 0.02)$ eV. This 
leads to the effective Majorana mass  of the electron neutrino 
\be
m_{ee}= m_1 \cos 2\theta_{\odot}~, 
\label{mee}
\ee
which can be  as large as $\sim O(10^{-2})$  eV if the solar mixing is not
to close to maximal one. 
(ii) The scheme predicts rather large mixing of $\nu_{\tau}$ and 
$\nu_s$ in the heaviest eigenstate (\ref{tau-4}), so that oscillation   
$\nu_{\tau}\leftrightarrow\nu_{s}$ drived by $\Delta m^2\sim \Delta m^2_{LSND}$
should be observed.

\section{$\nu_s$ -mixing and  the see-saw mechanism}

Let us show that $\nu_s$ -mixing considered in the previous sections is
consistent with the 
see-saw mechanism for the mass generation of the  active neutrinos. 

There are several possibilities to couple the sterile neutrinos with the
active neutrinos. 
The simplest one is the direct coupling  
through the 
Yukawa interaction $h_s \bar \nu_L \nu_s H$~, where
$H$ is the Higgs doublet. In this case, however,  one needs a very 
tiny coupling constant, 
$h_s \approx \epsilon m_{ss}/\langle H \rangle \approx 5\cdot 10^{-13}$,
and an additional mechanism is required to explain such 
a small value. 

Another possibility is the indirect mixing via coupling of 
$\nu_s$ with  heavy right handed  neutrinos $N_R$~\cite{ref13}.   
Let us consider the following mass terms
 \[
  -{\cal L} = \bar\nu_L m_D N_R + N_R^T M_R N_R + 
\bar\nu_s \mu^\dagger N_R + m_s \nu_s^T  \nu_s + h.c.~, 
 \]
where $\mu$ is the mass parameter of the  
$\nu_s$ and $N_R$ mixing. 
After decoupling of the right handed neutrinos the matrix $\hat M$~  
is produced with the folowing elements 
\ba
&& \hat m_f = - m_D  M_R^{-1} m_D^\dagger~; \quad
 \epsilon_{\alpha} m_{ss}= -
[m_D^\dagger M_R^{-1}]_{\alpha\beta}~\mu_\beta~; \nonumber\\
&& m_{ss}= m_s - \mu^\dagger M_R^{-1} \mu ~.
 \label{eq15}
 \ea 
The values of $\epsilon_\alpha$ are determined  by
$\mu_\alpha$, with  a scaling factor $\langle H\rangle/ M$. 
In order for $\epsilon_{\alpha}$ to be small enough we 
need $\mu$ to be at the level of the $m_D$ scale. Also  the  
corrections to $m_{ss}$ should be sufficiently 
suppressed by the right handed mass scale. 

The matrix $\hat M$ formed at the scale 
$M_R$ will be renormalized by the radiative corrections. 
In particular, the gauge (weak) interactions will renormalize differently 
the flavor submatrix and the mixing mass vector: 
\be
\hat m_f \rightarrow  r_g \hat m_f, ~~~ 
\epsilon_{\alpha} m_{ss}
 \rightarrow \sqrt{r_g} \epsilon_{\alpha} m_{ss}~, 
\ee
where $r_g \sim (1 - 2)$ is the renormalization factor which 
is model dependent. Clearly,  $m_{ss}$ is unchanged. 
Yukawa couplings can produce rather  small  
renormalization effect (mainly of the mass terms $M_{a \tau}$, 
where $a = e, \mu, \tau, s$). 
The renormalization should be taken into account when fixing 
parameters at the high (probably GUT) mass scale.

In the non-decoupling 
limit, $m_{ss}$ should be small. Such a small mass can be explained by
the see-saw mechanism in the ``sterile sector". 
The scale can be protected by  some
additional $U(1)$ symmetry, which is broken at  $\sim
(1-100)$ TeV scale~\cite{langacker}. In the context of the supersymmetric 
theories a smallness of  $m_{ss}$ can be related to physics of SUSY
breaking~\cite{benakli}.

\section{Discussion and Conclusion}

We have shown that mixing between active neutrinos can be enhanced by  
the presence  of  sterile neutrino. 
In particular, large or even maximal  mixing can 
be induced in the  flavor mass matrix with original strong mass hierarchy 
and small mixing. 

A  simultaneous explanation of the atmospheric and solar neutrino data 
requires however  certain structure of the original flavor mass matrix  (or the
introduction of  more than one sterile neutrino).  For  hierarchical flavor
mass matrix of general form this is impossible.   Indeed, mixing with just one
sterile neutrino produces three independent  mass terms,  whereas four
oscillations parameters (two $\Delta m^2$ and two mixing  angles)  should be
explained.

We have considered several cases which differ by the scale of  elements in
the original flavor mass matrix (larger, of the order, or smaller than the 
atmospheric  neutrino scale). For each case we have found the relevant
structure of the  flavor mass matrix. These structures look rather  plausible
from the point  of view of theories with flavor symmetries.  
Clearly, more freedom appears if there are  additional  sterile
neutrinos.  

Near maximal mixing appears as an interplay of the  original flavor  matrix and
the mass matrix induced by mixing  with sterile neutrino. If  the origin of the
mixing terms ($\epsilon$)  is unrelated to the origin of the flavor mass matrix,
then near maximal  mixing appears accidental.  In fact, in the case of mixing
via right handed neutrinos (37) the mixing terms may not be completely
independent of the elements of  flavor matrix:  both of them are proportional
to the product $m_D M_R^{-1}$. The required  relations  between  the 
$\epsilon$ - terms and the flavor matrix at some high scale may be  rather
non-trivial in view of renormalization effects.  

In our framework there is no simple symmetry which leads to maximal mixing. 
However it is not excluded  that  the required relation between $\epsilon$ 
terms  and the flavor matrix  appears in the underlying theory of flavor.  An
exception is the case IIIC, where both masses and mixing relevant for the 
atmospheric neutrino oscillations are given by the   dominant contribution
from  the sterile to active neutrino mixings.  In this case  near maximal
flavor mixing can appear due to  certain symmetry (equality) of the
$\epsilon$-elements themselves. 

In this paper we have concentrated on the phenomenological aspects of the 
mechanism.   Depending on the mass of sterile neutrino, two limits have been
considered.  In the  decoupling limit, mixings of active and sterile neutrinos
are very small. They are negligible for the laboratory experiments.   The
effect of sterile neutrinos is just reduced to correction of the mass matrix of
the active neutrinos. However, the sterile neutrinos with masses in (1 - 100)
kev range  and even very small mixing  can have important implications for
astrophysics and cosmology. 

In the non-decoupling limit, large  lepton mixings is accompanied by an
observable active-sterile neutrino mixing. Moreover,  for $\hat m_f$ with
moderate hierarchy the (3+1) scheme of neutrino mass~\cite{olg} can be
realized  which accommodate oscillation solutions of  the atmospheric and solar
neutrino problems  and the LSND result. The generic prediction of such a scheme
is a large $\nu_e-\nu_s$ mixing which leads to observable 
$\nu_e\leftrightarrow\nu_s$ oscillations.

The suggested mechanism of enhancement of mixing can reconcile  large observed
mixings and the  hierarchical structure of the active neutrino mass matrix 
(having small mixings), which naturally follows from the see-saw mechanism in
the  context of the Grand Unified Theories. 

\vskip2em

Balaji wishes to thank the High Energy Theory group at The Abdus Salam ICTP
for local support and for the hospitality at the Theoretical Physics 
Division, University of Helsinki, where this work was completed.

\end{document}